\documentclass{KapProc} 

\usepackage{aabib99}
\usepackage{epsfig}
\newcommand{\jahre}{\ensuremath{\, \mathrm{yr}}}
\newcommand{\msun}{\ensuremath{\, {\rm M}_\odot}}
\newcommand{\mdot}{\ensuremath{\dot{M}}}
\newcommand{\mem}[1]{\ensuremath{\mathrm{ #1}}}
\newcommand{\etal}{et~al.\,}
\newcommand{\abb}[1]{Fig.\,\ref{#1}}
\newcommand{\lsun}{\ensuremath{\, {\rm L}_\odot}}
\newcommand{\cdr}{\ensuremath{^{13}\mem{C}}}
\newcommand{\hedr}{\ensuremath{^{3}\mem{He}}}
\newcommand{\n}{\ensuremath{\mem{n}}}
\newcommand{\ose}{\ensuremath{^{16}\mem{O}}}

\let\footnote\savefootnote
\let\footnotetext\savefootnotetext 
\let\footnoterule\savefootnoterule 

\kluwerbib

\startauthorindex

\begin{document}
\include{abbrev}
\include{abbr-engl}



\articletitle[]{Stellar evolution and nucleosynthesis of Post-AGB Stars}

\chaptitlerunninghead{}

\author{Falk Herwig}
\affil{Institut f\"ur Physik, Abteilung Astrophysik, Universit\"at Potsdam, Potsdam, Germany}
\affil{Institute for Physics and Astronomy, University of Victoria, Victoria BC, Canada}
\email{fherwig@uvastro.phys.uvic.ca}

\begin{abstract}
I discuss recent new models of post-Asymptotic Giant Branch 
stellar evolution. These models aim to clarify the
evolutionary origin and status of a variety of hydrogen-deficient
post-AGB stars such as central stars of planetary nebulae of Wolf-Rayet
spectral type, PG1159 stars or Sakurai's object. Starting with AGB
models with   
overshoot such stars can evolve through one of four distinct channels. Each
of these channels has typical abundance patterns depending on the
relative timing of the departure from the AGB and the occurrence of the last
thermal pulse. I discuss the
responsible mechanisms and observational counterparts.
\end{abstract}


\section{Introduction}
Nuclear burning and mixing processes in post-Asymptotic Giant Branch
(pAGB) stars are closely
related to the final thermal pulse \cite{iben:83a,iben:95}. 
The pAGB stage will proceed undisturbed if the
final thermal pulse (TP) occurs well before the departure from the
AGB.  However, if the final TP
occurs immediately before the 
departure from the AGB (AFTP case) or during the early (LTP case) or the late
post-AGB phase (VLTP case), then internal mixing and possibly as well
nuclear burning    
processes will lead to distinctive alterations of the surface
abundances \cite{herwig:00a,bloecker:00a}. For example, the
hydrogen-deficiency of central stars of  
planetary nebulae of Wolf-Rayet  
spectral type ([WC]-CSPN) 
is the result of such an abundance alteration as concluded
from their large helium, carbon and oxygen
abundance and the lack of hydrogen in their spectra (e.g.\ Koesterke
and Hamann, 1997). \nocite{koesterke:97b}

The final TP affects  other elements as well. This  is 
supported by the weird abundance pattern of Sakurai's
object  which  apparently is a
born-again AGB object \cite{asplund:99a}. It shows a real-time
evolution of lithium, 
hydrogen, $s$-process elements  
and others.
Hybrid objects -
with an intermediate hydrogen abundance between 
 [WC]-CSPN and H-normal stars - are as well associated with 
a non-standard pAGB evolution \cite{napiwotzki:91}.

The TP is a thermonuclear instability of the He-shell and
accordingly a TP can occur until  He-burning ends at the latest at the
pre-WD stage of pAGB evolution. On the AGB TPs are recurring mainly
independently of mass loss on a timescale of a $10^4$ to $10^5\jahre$. 
The AGB
evolution is ended randomly by mass loss which strips the envelope
off the core. If the envelope mass decreases below a critical value
of about $0.001\msun$ contraction sets in and the star will begin
its evolution along the horizontal part of pAGB evolution in the
HRD. The last TP which
occurs in the lifetime of a star may be past  a few thousand years
when the AGB star starts contraction to 
evolve into a central star. In that case this last TP does not interfere with
an undisturbed and hydrogen-rich pAGB evolution finally leading to a white
dwarf of spectral type DA. 

However, if by coincidence the last TP occurs either
immediately before or within a certain time interval after the
transition towards the central star phase then a severely different
evolution must be expected. In the following sections I will discuss
in turn the three channels of evolution in which the final TP
interferes with an undisturbed pAGB evolution. 

\section{The AGB Final Thermal Pulse}

We found in
our calculations of AGB models with overshoot that dredge-up is much
more efficient if overshoot is not only applied to the bottom of the
envelope convection zone but as well  to the He-flash
convection zone in the intershell during the TP
\cite{herwig:97,herwig:99a}. 
While the third dredge-up is usually difficult to find in stellar
models of low core mass and low envelope mass we did find dredge-up
after the last TP still on the AGB at a surprisingly low envelope mass
of $4\cdot 
10^{-3}\msun$. This dredge-up was found under the assumption of an
increased efficiency of convective mixing ($\alpha_\mem{MLT}=3.0$) and
overshoot with an efficiency parameter of $f=0.016$. The amount of
dredged-up material was of the order of $3\cdot 10^{-3}\msun$. 

In the AFTP case the final TP coincidently occurs when the
envelope mass is already very close to the critical envelope mass
which marks the departure of the star from the AGB regime. The
remaining envelope mass and the amount of dredged-up mass are of
comparable quantity and therefore the surface abundance change due to
dredge-up is considerable. If the star is O-rich (C/O<1) before the
final TP it will turn into a carbon star during the following  final
dredge-up episode, just at the onset of the proto-planetary transition
phase. This carbon star is different from the C-stars known on
the AGB in that its C abundance is much larger. In fact, after
an AFTP dredge-up episode as described above the emerging central star 
is hydrogen-deficient. The extent of H-deficiency depends on the
amount of envelope mass at the final TP and the efficiency
of following dredge-up.
As an example, the two AFTP calculations by Herwig
\cite*{herwig:00a} produced central stars with mass fractions of
(H/He/C/O) of (0.55,0.31,0.07,0.04) and (0.17,0.33,0.32,0.15).

Although the current models of the AFTP have
systematically larger H-abundances than found in the majority of [WC]
CSPN or PG1159 stars
this evolutionary scenario has otherwise observational support. 
Planetary nebulae (PNe) of [WC]-type central stars
are on average not older than PNe surrounding chemically normal
central stars \cite{gorny:95}. This indicates
that these objects are on their first evolutionary
departure from the AGB. The alternative evolutionary scenario described
below, in particular the very late TP, would rather
require H-deficient central stars to have old PNe since they are
supposed to be on their second departure from the AGB following a
born-again evolution which might last more than $10^4\jahre$. Thus,
the AFTP scenario naturally resolves the time scale problem of
PNe around H-deficient CSPN. 

Recent  ISO observations support the AFTP
scenario as well \cite{waters:98,cohen:99}. Surprisingly, both oxygen and  
carbon rich dust has been found around [WC]-late ([WC-L])  CSPN
(called here the \emph{C+O dust feature}). C-rich dust 
(PAH = Polycyclic Aromatic Hydrocarbon) and O-rich dust
(crystalline silicates) usually exclude each other because of the initial 
formation of CO molecules. Consequently the simultaneous presence of O-
and C-rich dust must be the result of a recent change of surface abundance
pattern in the mass losing star. 

The AFTP scenario does 
provide such a sudden abundance change immediately before the onset
of the pAGB phase. In this picture the O-rich AGB star evolves
towards the final TP with presumably high mass loss and
experiences efficient dredge-up of intershell material which has a C/O
ratio of about $3\dots 4$. Instantly the star will become 
C-rich. Possibly the now C-rich surface composition will once more
enhance mass loss and accelerate the evolution towards the pAGB
phase. The timescale of the abundance change and the following
transition evolution off the AGB are around $1000\jahre$, in agreement with
the timescale inferred for the abundance change from the ISO observations
of the circumstellar dustshells.

If \emph{all} [WC-L] CSPN show the 
signature of 
simultaneous presence of O- and C-rich dust (see the
contribution by  Hony \etal, these 
proceedings) we may assume that \emph{all} [WC-L] CSPN are 
originating from an AFTP because this is currently the only scenario
we know for the C+O dust phenomenon. This raises the question why there are  no
[WC-L] type descendants of the LTP or VLTP evolution (see below)
in their second departure from the AGB. This would 
imply that a star like 
Sakurai's object does not show up as a [WC-L] type CSPN, possibly
because these objects camouflage themself by ejecting puffs of dust as 
suggested by another born-again candidate \mbox{V\,605 Aql}
\cite{clayton:97}. However, there is at least one [WC-L] star, \mbox{V\,348
Sgr}, which is likely to be a born-again object because of its old PNe 
\cite{pollacco:90} and thus, no O-rich dust signatures should be
detected for this object.

There might be as well another clue for the progenitors of [WC-L]
CSPN. If all or at least the overwhelming majority of
\mbox{[WC-L]} show the C+O dust feature then an AFTP 
predominantly occurs in O-rich AGB stars (S- and M-giants).
An AFTP event 
occurring in a C-giant could not  host crystalline silicates in its
dustshell.
Possibly, the mass loss of S- and M-giants depends more strongly than
that of C-giants
on the stellar parameter variation during a 
TP and thereby correlates the occurence of the final
TP with the 
immediately following departure from the AGB. Maybe the
C-giants  have a low fraction of immediate departures from the AGB
after a final TP because their mass loss is more evenly 
distributed over the TP.

\section{The Late Thermal Pulse}
\label{sec:ltp}
While in the AFTP case the final TP on the AGB is immediately followed 
by the transition of the star towards the pAGB phase the order of
events is reversed in the LTP case. The departure from the AGB begins
up to about $5000\jahre$ before the last TP. Accordingly the star will 
be disturbed by a TP during the pAGB evolution which leads to a
born-again evolution back to the AGB. This evolutionary scenario is
closely related to the VLTP described below with one main
distinction. In the LTP case hydrogen burning has not yet ceased at the
time of the TP. As a consequence the star will not experience the
violent nuclear processes induced by the ingestion of the H-rich
envelope material into the He-flash convection zone as found for
VLTP models (see below). Instead,  Bl\"ocker \cite*{bloecker:00a}
and Herwig \cite*{herwig:00a} found that  dredge-up during the 
born-again AGB stage after the LTP can lead to abundance alterations
which are in accordance with abundances of [WC]-CSPN as well. If
the time  
interval between the departure from the AGB and the TP is small this
scenario could even account for the normal ages of PNe around
[WC]-CSPN. Moreover this evolutionary channel more naturally leads to
the very low H-abundances actually observed in the 
[WC]-CSPN. The envelope mass is by definition of the order of
$10^{-4}\msun$ when the star has entered the CSPN stage and the TP
occurs. In the LTP calculation by Herwig \cite*{herwig:00a} the amount 
of dredged-up material is of the order of a few $10^{-3}\msun$ and the 
resulting surface mass fractions are
(H/He/C/O)=(0.02,0.37,0.40,0.18). 

Taking into account the comparatively large mass loss of [WC]-CSPN of $\mdot
\sim 10^{-6} \msun$ \cite{koesterke:97b}, it is conceivable that a pAGB 
star after a LTP born-again event loses its entire top layer during
the second pAGB evolution. This top layer represents the
region well mixed during the dredge-up after the LTP and consists of
the just mentioned amount of  a few $10^{-3}\msun$. If this stripping of the
outer dredge-up layer is possible for a LTP descendant star then the
LTP evolution could even yield non-DA WDs which do not contain any hydrogen 
at all.

One of the unsolved problems with this scenario is the description of
mass loss during the second, born-again AGB phase when the abundance
suddenly and dramatically changes due to dredge-up. The large
amounts of C and O brought to the surface could initiate a
rapid increase in dust formation and thereby mass loss.

\section{The Very Late Thermal Pulse}
\label{sec:vltp}
As in the LTP case the final TP occurs after the  departure from the
AGB. However, here the time interval between the AGB 
departure and the TP is longer, typically exceeding $\sim 5000
\jahre$. At this time  hydrogen burning has already
stopped and the entire envelope material will be included in the
convective He-burning region (see Herwig \etal \cite*{herwig:99c} and
Herwig \cite*{herwig:00a} for details and references). The star takes
a deep loop through the HRD in order to return to the AGB again,
similar to the LTP case.  While the VLTP model abundances agree with the
observations of [WC]-CSPN, the PNe timescale problem is 
most severe for this case. Before the potential of the LTP and the AFTP to
provide scenarios for the origin of the H-deficient [WC]-CSPN and
PG1159   was realized the VLTP was thought to be the only possible
evolutionary origin of these objects \cite{iben:95b}. 

The VLTP provides unique conditions  of convective nucleosynthesis
originating from the 
ingestion of protons from the unprocessed envelope into the He-flash
convection zone. The nuclear burning and mixing occur both on the 
time scale of about one hour once the protons have reached sufficient
temperature while being transported inward by convection. Temporarily
and locally the luminosity of hydrogen 
burning exceeds the peak-flash He-burning luminosity and the He-flash
convection zone is fragmented. 
\cite{asplund:99a}.
\begin{figure}[t]
\epsfxsize=8.8cm
\epsfbox{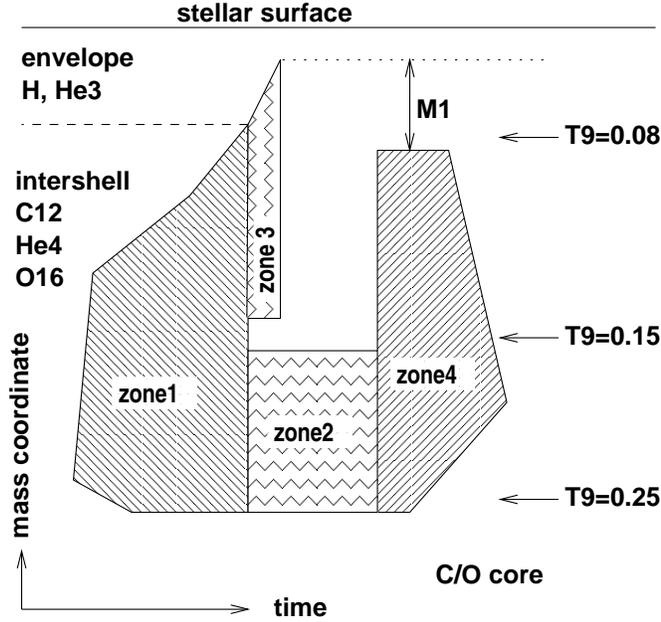}
\caption{ \label{fig:1} 
Schematic (not to scale) of the time-evolution of convection zones in the top
$0.01\msun$ of a post-AGB star of $0.6\msun$ during a VLTP. All shaded
areas indicate convectively unstable zones. The two solid horizontal
lines in the upper part of the diagram indicate the stellar surface
and the mass coordinate of the envelope-intershell (=core) transition.}
\end{figure}

The nucleosynthesis during  the VLTP is highly dependent on the
details of convective mixing because here both processes are closely
related. While the protons from the envelope enter a hotter
environment on their way down into the He-flash convection zone their
nuclear burning time scale diminishes. At the position where it matches 
the convective turnover time scale the greatest part of the
hydrogen-burning luminosity will be created. Therefore the position in 
the star where the peak H-burning luminosity can be found will depend
both on the assumptions  of convective mixing efficiency and
the numerical treatment of the simultaneous burning and mixing. 

For the  calculation by Herwig \etal\cite*{herwig:99c} we developed a fully
coupled numerical scheme for convective nucleosynthesis. The equations 
of material transport (one diffusion-like equation for each isotope)
and the nuclear network equations at each depth mass grid are solved
altogether fully implicit in one scheme \cite{herwig:00a}. This
treatment returns 
consistent abundance profiles within the convective region. Thus, the
energy generation rates calculated as a function of position in the
stellar burning region reflect the rapid consumption and simultaneous
convective mixing of protons.

The evolution of the pulse-driven convection
zone during a VLTP model sequence is sketched in \abb{fig:1}. It
shows the sequence of events found in the models of Herwig
\etal\cite*{herwig:99c}.  The convective instability can be divided in 
four different regions. Zone 1 represents the onset of the
He-flash. During this first phase the He-burning luminosity increases
by several orders of magnitude. It is essentially the same evolution as
during the onset of any AGB TP or the LTP. The difference
occurs when the upper boundary of the He-flash convection zone
approaches the H-rich envelope (second horizontal line from top in
\abb{fig:1}). In the VLTP case the convective instability reaches out
of the intershell and spreads into the H-rich envelope as represented by 
zone 3 in \abb{fig:1}. This is not possible in the LTP case or during 
typical AGB TP because there the H-burning shell is still
active at the time of the TP \cite{iben:76}. The fresh
nuclear fuel which is ingested into the deeper and hotter region
burns on ever shorter timescales while at the bottom of zone 2 the
unstable He-burning  continues mainly undisturbed. 

Due to the locally large energy generation of up to $10^8\lsun$ by H-burning
a small radiative layer between zone 2 and 3 establishes. This
radiative layer affects the permeability for particle transport 
from zone 2 to 3.  If overshoot is very efficient
 a considerable fraction of freshly produced
particles, e.g.\ \cdr\ will be burned at the bottom of zone 3 instead
of the bottom of zone 2. Thus, the assumed overshoot efficiency has an 
important influence on the model predictions. Moreover the position of 
the split between zone 2 and 3 depends on the assumption of convective 
efficiency. For a larger convective efficiency the mass coordinate of
peak H-burning luminosity and thereby of the split shifts inwards and
accordingly the maximum temperature ruling the nucleosynthesis in zone 
2 increases.

The additional energy supply by proton captures causes zone 2 to
stretch out and engulf the entire envelope. The moment of abundance
change at the surface due to the mixing of zone 2 depends on the
initial conditions of the VLTP. On one extreme is the immediate
abundance change after the VLTP while the star still resides in the
blue and faint part of the born-again loop in the HRD. This will
only occur for the most vigorous cases of proton burning during the
VLTP. In other cases a tiny layer of original envelope material of
less than $10^{-6}\msun$ may cover the changed abundances close to the 
surface until during the evolution back to the AGB either mass loss or 
the onset of convection will reveal the H-poor abundance.

In any case, zone 3 is only short-lived due to the restricted amount of 
hydrogen available in the envelope. If that amount is consumed the
related convective region will die away. After some
cooling of that layer the original He-flash convection zone resumes
control (zone 4). Note, that the most outward reach of zone 4 is
smaller than that of zone 3. In the layer M1 swept over by zone 3 but
not  by zone 4 an element distribution formed only in zone 3  can be
preserved close enough to the stellar surface in order to show up
later on. It is here where we suspect lithium to form according to the
mechanism of \emph{hot hydrogen-deficient \hedr\ burning}
\cite{herwig:00f}. This lithium could explain the
observed abundance  in Sakurai's 
object \cite{asplund:99a} which is believed to be a VLTP star caught in its
born-again evolution.

The VLTP does also provide a rich neutron capture nucleosynthesis
initiated by the production of \cdr\ in zone 3 \cite{malaney:86}. The
neutrons are partly released immediately by the well known
$\cdr(\alpha,\n)\ose$ reaction. Alternatively \cdr\ is stored in the
region covered by zone 3 until the He-flash convection zone recovers
from the disturbation of H-burning. Then \cdr\  burns at higher
temperatures in zone 4. Future models of the $s$-process during the
VLTP should predict a distinctive heavy-element signature to be
compared with that of Sakurai's object.
 

\begin{acknowledgments}
This work has been supported by the \emph{Deut\-sche
  For\-schungs\-ge\-mein\-schaft, DFG\/} (La\,587/16).  I would like
to thank T.\ Bl\"ocker, W.-R.\ Hamann, L.\ Koesterke, N.\ Langer, D.\
Sch\"onberner and K.\ Werner for very useful discussions. 
\end{acknowledgments}


\begin{chapthebibliography}{1}

\bibitem[\protect\astroncite{{Asplund} et~al.}{1999}]{asplund:99a}
{Asplund}, M., {Lambert}, D.~L., {Kipper}, T., {Pollacco}, D., and {Shetrone},
  M.~D., 1999,
\newblock {A\&A} {343}, 507

\bibitem[\protect\astroncite{Bl\"ocker}{2000}]{bloecker:00a}
Bl\"ocker, T., 2000,
\newblock in T. Bl\"ocker, R. Waters, and B. Zijlstra (eds.), {Low mass
  Wolf-Rayet Stars: origin and evolution}, Ap\&SS,
\newblock in press

\bibitem[\protect\astroncite{{Clayton} and {de Marco}}{1997}]{clayton:97}
{Clayton}, G.~C. and {de Marco}, O., 1997,
\newblock {AJ} {114}, 2679

\bibitem[\protect\astroncite{{Cohen} et~al.}{1999}]{cohen:99}
{Cohen}, M., {Barlow}, M.~J., {Sylvester}, R.~J., {Liu}, X.~., {Cox}, P.,
  {Lim}, T., {Schmitt}, B., and {Speck}, A.~K., 1999,
\newblock {ApJ Lett.} {513}, L135

\bibitem[\protect\astroncite{{Gorny} and {Stasinska}}{1995}]{gorny:95}
{Gorny}, S.~K. and {Stasinska}, G., 1995,
\newblock {A\&A} {303}, 893

\bibitem[\protect\astroncite{Herwig}{2000a}]{herwig:99a}
Herwig, F., 2000a,
\newblock {A\&A} {360}, 952

\bibitem[\protect\astroncite{Herwig}{2000b}]{herwig:00a}
Herwig, F., 2000b,
\newblock in T. Bl\"ocker, R. Waters, and B. Zijlstra (eds.), {Low mass
  Wolf-Rayet Stars: origin and evolution}, Ap\&SS,
\newblock in press, astro-ph/9912353

\bibitem[\protect\astroncite{Herwig et~al.}{1999}]{herwig:99c}
Herwig, F., Bl\"ocker, T., Langer, N., and Driebe, T., 1999,
\newblock {A\&A} {349}, L5

\bibitem[\protect\astroncite{Herwig et~al.}{1997}]{herwig:97}
Herwig, F., Bl\"ocker, T., Sch\"onberner, D., and {El Eid}, M.~F., 1997,
\newblock {A\&A} {324}, L81

\bibitem[\protect\astroncite{Herwig and Langer}{2000}]{herwig:00f}
Herwig, F. and Langer, N., 2000,
\newblock {Nucl. Phys. A},
\newblock in press, astro-ph/0010120

\bibitem[\protect\astroncite{Iben}{1976}]{iben:76}
Iben, Jr., I., 1976,
\newblock {ApJ} {208}, 165

\bibitem[\protect\astroncite{Iben}{1995}]{iben:95}
Iben, Jr., I., 1995,
\newblock {Physics Reports} {250}, 1

\bibitem[\protect\astroncite{Iben et~al.}{1983}]{iben:83a}
Iben, Jr., I., Kaler, J.~B., Truran, J.~W., and Renzini, A., 1983,
\newblock {ApJ} {264}, 605

\bibitem[\protect\astroncite{Iben and MacDonald}{1995}]{iben:95b}
Iben, Jr., I. and MacDonald, J., 1995,
\newblock in D. Koester and K. Werner (eds.), {White Dwarfs}, No. 443 in LNP,
  p.~48, Springer, Heidelberg

\bibitem[\protect\astroncite{{Koesterke} and {Hamann}}{1997}]{koesterke:97b}
{Koesterke}, L. and {Hamann}, W.~R., 1997,
\newblock {A\&A} {320}, 91

\bibitem[\protect\astroncite{{Malaney}}{1986}]{malaney:86}
{Malaney}, R.~A., 1986,
\newblock {MNRAS} {223}, 683

\bibitem[\protect\astroncite{Napiwotzki et~al.}{1991}]{napiwotzki:91}
Napiwotzki, R., Sch\"onberner, D., and Weidemann, V., 1991,
\newblock {A\&A} {243}, L5

\bibitem[\protect\astroncite{{Pollacco} et~al.}{1990}]{pollacco:90}
{Pollacco}, D.~L., {Hill}, P.~W., and {Tadhunter}, C.~N., 1990,
\newblock {MNRAS} {245}, 204

\bibitem[\protect\astroncite{Waters et~al.}{1998}]{waters:98}
Waters, L. B. F.~M., Beintema, D.~A., Zijlstra, A.~A., and {et al.}, 1998,
\newblock {A\&A} {331}, L61


\end{chapthebibliography}

\end{document}